\documentclass[proof]{WileyASNA-v1}
\usepackage{bm}
\usepackage[utf8]{inputenc}

\articletype{Proceedings}%

\received{31 December 2020}
\revised{31 December 2020}
\accepted{31 December 2020}

\begin{document}

\title{Pulsar tests of the graviton mass\protect\thanks{Invited parallel
talk at the 9th International Workshop on Astronomy and Relativistic
Astrophysics, 6--12 September, 2020.}}

\author[1,2]{Lijing Shao}

\authormark{LIJING SHAO}

\address[1]{\orgdiv{Kavli Institute for Astronomy and Astrophysics},
\orgname{Peking University}, \orgaddress{\state{Beijing 100871},
\country{China}}}

\address[2]{\orgdiv{National Astronomical Observatories}, \orgname{Chinese
Academy of Sciences}, \orgaddress{\state{Beijing 100012}, \country{China}}}

\corres{*\email{lshao@pku.edu.cn}}
\abstract{In Einstein's general relativity (GR), gravity is described by a
massless spin-2 metric field, and the extension of GR to include a mass
term for the graviton has profound implication for gravitation and
cosmology. Besides the gravity experiments carried out in the Solar System
and those recently with gravitational waves (GWs), pulsar timing observations
provide a complementary means to test the masslessness of graviton. In this
contribution, I overview three methods in probing the mass of graviton from
precision timing of binary pulsars via the modified gravitational radiation
(hence, the observed damping rate of the orbital period), as well as from
the pulsar-timing-array (PTA) experiments via the modified Hellings-Downs
angular-correlation curve. These tests probe different aspects of
gravitation in its kinematics and dynamics, complementing tests of other
kinds and providing valuable information to the fundamental theory of
gravity.}


\jnlcitation{\cname{%
\author{Shao L.}} (\cyear{2020}), 
\ctitle{Pulsar tests of the graviton mass}, \cjournal{Astron. Nachrichten},
\cvol{2020;ZZZ:XX--YY}.}


\maketitle

\footnotetext{\textbf{Abbreviations:} GR, general relativity; GW,
gravitational wave; PTA, pulsar timing array}

\section{Introduction}\label{sec:intro}

Modern physics is built upon two fundamental pillars, the Standard Model of
particle physics and the General Relativity (GR) of gravitation. The former
is nicely spoken in the language of quantum field theory \citep[see
e.g.,][]{Weinberg:1995mt}, while the latter is communicated in the
narrative of differential geometry \citep[see e.g.,][]{Misner:1974qy}.
Together, they account for the four fundamental forces in the Nature,
namely, the electromagnetic force, the strong force, the weak force, and
the gravitational force. The incentive to unify quantum field theory and GR
into a fundamental theory, the often so-called quantum gravity, is the
driving motivation for most of the past decades' investigation in
fundamental physics.

Among the four fundamental forces, gravity is quite unique from the other
three. It might be holding the key towards new physics beyond what we
currently know of. Therefore, theoretical and experimental studies in
gravitational physics have been constantly carried out with great
enthusiasm \citep{Will:2018bme}. In Einstein's theory, gravity is described
by a massless spin-2 metric field, and the extension to include a mass term
for the graviton has profound implication for gravitation and cosmology
\citep{deRham:2014zqa}. In this contribution, we focus on the possibility
to include a mass term, and overview the methods to bound it utilizing the
precision timing of radio pulsars, which is one of the most precise
strong-field experiments in the field of experimental gravity
\citep{Taylor:1994zz, Wex:2014nva, Kramer:2016kwa, Shao:2016ezh}.

From the theoretical side, the mass of graviton was firstly perceived by
\citet{Fierz:1939ix} in late 1930s. After fighting for decades with
pathologies like the van Dam-Veltman-Zakharov discontinuity and
Boulware-Deser ghosts, some version of healthy massive gravity theory was
developed by the gravity community \citep[see e.g.,][for a
review]{deRham:2014zqa}. From the experimental side, various observations
were used to bound the mass of graviton, including---just to name a
few---the propagation of gravitational waves (GWs) \citep{Will:1997bb} and
the perihelion advance rate of planets in the Solar System
\citep{Will:2018gku}. Interested readers are referred to
\citet{deRham:2016nuf} for a recent comprehensive review.

In this short contribution, I will overview some investigation using the
precision timing of radio pulsars in bounding the mass of graviton. In
particular, I will cover the following studies.
\begin{enumerate}[(I)]
  \item The Finn-Suttom method \citep{Finn:2001qi} was used to bound the
  graviton mass to be $m_g \lesssim \mbox{a few} \times 10^{-21} \, {\rm
  eV/}c^2$ \citep{Miao:2019nhf} in a dynamic regime for a Fierz-Pauli-like
  gravity action.
  \item The scheme developed by \citet{deRham:2012fw} was used to bound the
  graviton mass to be $m_g \lesssim \mbox{a few} \times 10^{-28} \, {\rm
  eV/}c^2$ \citep{Shao:2020fka} in the cubic Galileon theory.
  \item In the near future, the Hellings-Downs angular-correlation curve
  will be used to bound the graviton mass to be $m_g \lesssim \mbox{a few}
  \times 10^{-22} \, {\rm eV/}c^2$ with pulsar timing arrays
  \citep[PTAs;][]{Lee:2010cg, Lee:2014awa}.
\end{enumerate}
These tests are of different nature and when performing comparison among
them, I strongly argue to specify the context concerning the underlying
physics and assumptions. Unless otherwise stated, I use units where $G=c=1$
in the manuscript.

\section{The Finn-Sutton test}\label{sec:finn}

\citet{Finn:2001qi} considered a phenomenological Fierz-Pauli-like action
for linearized gravity with a mass term for the transverse tensor modes,
\begin{align}
  S \sim & \int \mathrm{d}^{4} x\bigg[\partial_{\lambda} h_{\mu
  \nu} \partial^{\lambda} h^{\mu \nu}-2 \partial^{\nu} h_{\mu \nu}
  \partial_{\lambda} h^{\mu \lambda}+2 \partial^{\nu} h_{\mu \nu}
  \partial^{\mu} h \nonumber \\
  & -\partial^{\mu} h \partial_{\mu} h-32 \pi h_{\mu \nu} T^{\mu
  \nu}+m_{g}^{2}\left(h_{\mu \nu} h^{\mu \nu}-\frac{1}{2}
  h^{2}\right)\bigg] \,. \label{eq:Finn:Sutton}
\end{align}
The first few terms are from the linearized version of GR
\citep{Misner:1974qy}, while the last term is the mass term of specific
interests here. The mass term is unique when requiring, (i) a standard
Klein-Gorden-like wave equation for $h_{\mu\nu}$, and (ii) a recovery of GR
if the graviton mass $m_g$ goes to zero. Though the simple model
(\ref{eq:Finn:Sutton}) contains ghosts and instabilities
\citep{Boulware:1973my}, it is nevertheless a valuable strawman target to
study massive gravity as an illustration. Nevertheless, it should not be
taken as a full and sophisticatedly designed theory at the end.

Assuming slow motion for a Keplerian binary orbit as a reasonable
approximation for binary pulsars, \citet{Finn:2001qi} showed that there is
a correction to the orbital decay rate as predicted by GR,
\begin{equation}
  \frac{\dot{P}_{b}-\dot{P}_{b}^{\mathrm{GR}}}{\dot{P}_{b}^{\mathrm{GR}}}
  =\frac{5}{24} \frac{\left(1-e^{2}\right)^{3} }{1+\frac{73}{24}
  e^{2}+\frac{37}{96} e^{4} } \left(\frac{P_{b}}{2 \pi \hbar}\right)^{2}
  m_{g}^{2} \,, \label{eq:Pbdot:diff}
\end{equation}
where $P_b$ is the orbital period, and $e$ is the orbital eccentricity. The
orbital decay rate in GR, due to the emission of GWs, is,
\begin{equation}
  \dot{P}_{b}^{\mathrm{GR}}=-\frac{192 \pi }{5} \frac{1+\frac{73}{24}
  e^{2}+\frac{37}{96} e^{4}}{\left(1-e^{2}\right)^{7 / 2}} \left(\frac{2
  \pi}{P_{b}}\right)^{5 / 3} \frac{m_{1} m_{2}}{\left(m_{1}+m_{2}\right)^{1
  / 3}} \,, \label{eq:Pbdot:GR}
\end{equation}
for a binary of component masses $m_1$ and $m_2$ \citep{Peters:1963ux}.

For a handful of binary pulsars, the masses can be derived via measuring
the post-Keplerian parameters \citep{Damour:1991rd, Taylor:1994zz}, while
at sometimes, in combination with optical phase-resolved spectroscopic
observation of the companion \citep{Wex:2014nva, Ozel:2016oaf}. They can be
used to derive the theoretical orbital decay rate in GR via
Eq.~(\ref{eq:Pbdot:GR}). On the other hand, we can derive the value of
$\dot P_b$ directly from the pulsar timing data. However, this value,
$\dot{P}_{b}^{\mathrm{obs}}$, in general is contaminated by astrophysical
contribution of various sources \citep{Lorimer:2005misc}. One has to
subtract these contributions, in order to get the intrinsic orbital decay
rate
\begin{equation}
  \dot{P}_{b}^{\mathrm{intr}} = \dot{P}_{b}^{\mathrm{obs}} -
  \dot{P}_{b}^{\mathrm{acc}}-\dot{P}_{b}^{\mathrm{Shk}} \,,
  \label{eq:Pbdot:intr}
\end{equation}
where we have denoted the two most significant ones, namely the
contribution $\dot{P}_{b}^{\mathrm{acc}}$ caused by the difference of
accelerations of the binary pulsar and the barycenter of the Solar System
projected along the line of sight to the pulsar \citep{Damour:1990wz}, and
the ``Shklovskii'' contribution $\dot{P}_{b}^{\mathrm{Shk}}$ caused by the
relative kinematic motions of the binary pulsar with respect to the
barycenter of the Solar System \citep{Shklovskii:1970}.

After obtaining the intrinsic orbital decay rate, a meaningful bound on the
graviton mass can be derived via using Eq.~(\ref{eq:Pbdot:diff}).
\citet{Miao:2019nhf} carefully chose a few best timed binary pulsars, and
performed a thorough study. In their Bayesian analysis that combines all
these binary pulsars, they obtained,
\begin{equation}
  m_{g} < 5.2 \times 10^{-21} \, \mathrm{eV} / c^{2} \,, \quad \mbox{(90\%
  {C.L.})} \,. \label{eq:Finn:Sutton:limit}
\end{equation}
Though failing to compete with some other graviton mass bounds under
different contexts \citep{deRham:2016nuf}---like using the static Yukawa
potential or the modified dispersion relation of GWs---the bound in
Eq.~(\ref{eq:Finn:Sutton:limit}) is the currently best limit on the mass
term in action (\ref{eq:Finn:Sutton}) from binary pulsars in a dynamic
regime. It encodes two-body dynamics instead of pure kinematics or static
Yukawa-type suppression.

\section{The cubic-Galileon test}\label{sec:cubic}

The Lovelock theorem is a useful guide to classify modified gravity
theories \citep{Berti:2015itd}. According to it, in a 4-dimensional
spacetime the only diffeomorphism invariant, divergence-free, symmetric
rank-2 tensor, that is constructed solely from the metric and its
derivatives up to the second order, is the Einstein tensor plus a
cosmological term \citep{Lovelock:1972vz}. Consequently, unlike what is in
Eq.~(\ref{eq:Finn:Sutton}), a full massive gravity often introduces extra
scalar degrees of freedom \citep{deRham:2014zqa}. Salient features of
various massive gravity theories are captured by Galileon models, and here
we will discuss the cubic Galileon theory \citep{Luty:2003vm}, which is the
simplest one of them. The cubic Galileon is often taken as a proxy to all
of the Lorentz-invariant massive gravity models in some appropriate limits
\citep{Nicolis:2008in, deRham:2014zqa}.

For the orbital dynamics of binary pulsars, we consider the action
\citep{deRham:2012fw,Shao:2020fka},
\begin{align}
  S \sim & \int \mathrm{d}^{4} x\bigg[-\frac{1}{4} h^{\mu \nu}(\mathcal{E}
  h)_{\mu \nu}+\frac{h^{\mu \nu} T_{\mu \nu}}{2 M_{\mathrm{Pl}}} \nonumber
  \\
  & -\frac{3}{4}\left(\partial \pi\right)^{2}\left(1+\frac{1}{3
  \Lambda^{3}} \square \pi\right)+\frac{\pi T}{2
  M_{\mathrm{Pl}}}\bigg] \,, \label{eq:cubic}
\end{align}
where the first two terms are from the linearized GR, $\pi$ is the scalar
field with Galileon symmtry, and $\Lambda$ is the strong coupling energy
scale related to the graviton mass by $\Lambda^{3}=m_{g}^{2}
M_{\mathrm{Pl}}$ with $M_{\mathrm{Pl}}$ the Planck mass. The Vainshtein
screening radius $r_\star$ \citep{Vainshtein:1972sx} is given via
$r_\star^3 = {M}/{16 m_{g}^{2} M_{\mathrm{Pl}}^{2}}$.

\citet{deRham:2012fw} discovered that, in binary systems, the suppression
factor in the extra gravitational radiation channels due to the Galileon
mode is less than that in the static fifth force. For this reason, it is
interesting to check with the radiative tests in binary pulsars. These
authors worked out the explicit expressions for extra monopole, dipole, and
quadrupole radiations in the cubic Galileon theory, in addition to what is
predicted from GR in Eq.~(\ref{eq:Pbdot:GR}). These extra contributions are
proportional to $m_g$, instead of proportional to $m_g^2$ as in
Eq.~(\ref{eq:Pbdot:diff}) for the Fierz-Pauli-like theory; explicit
equations can be found in \citet{deRham:2012fw} and \citet{Shao:2020fka}.

Similarly to the linearized Fierz-Pauli theory, extra gravitational
radiations lead to a faster orbital period damping rate, by an amount of
\begin{equation}
  \dot P_b^\pi = \dot P_b^{\rm mono} + \dot P_b^{\rm dipo} + \dot P_b^{\rm
  quad} \,, \label{eq:Pbdot:Pi}
\end{equation}
which can be confronted with experiments via the measurement of the
intrinsic $\dot P_b^{\rm intr}$ in Eq.~(\ref{eq:Pbdot:intr}). The
dependence of the Galileon contributions on the orbital period, orbital
eccentricity, and binary component masses is complicated, and a full
discussion can be found in \citet{Shao:2020fka}.

\begin{figure}[t]
	\centerline{\includegraphics[width=85mm]{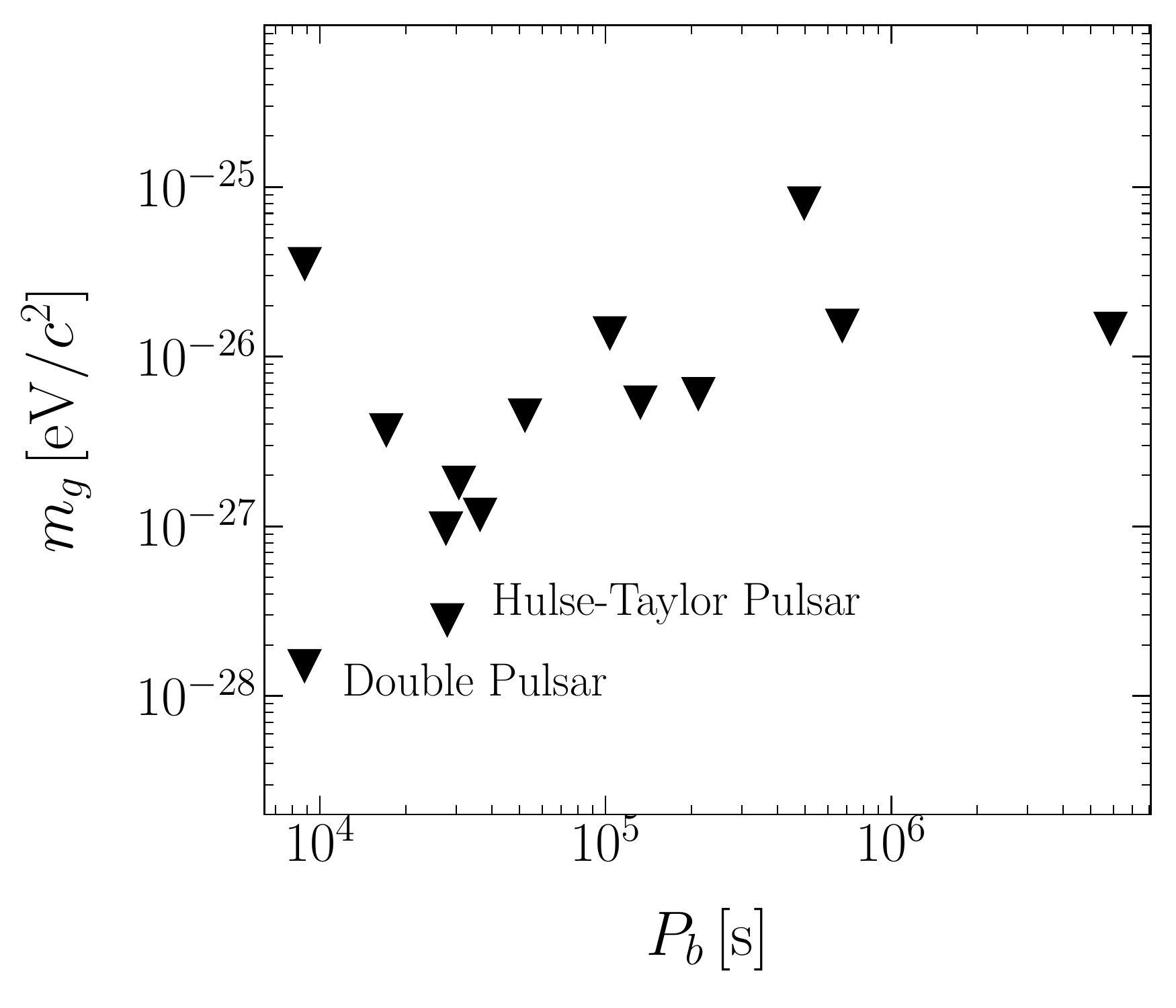}}
	\caption{Bounds on the graviton mass in the cubic Galileon theory from
	different binary pulsars versus the orbital period
	\citep{Shao:2020fka}. The best two pulsars in the test are denoted.
	\label{fig:bounds}}
\end{figure}

A dozen of well-timed binary pulsar systems were carefully chosen to
confront with Eq.~(\ref{eq:Pbdot:Pi}), including the famous Hulse-Taylor
pulsar PSR~B1913+16 \citep{Weisberg:2016jye} and the Double Pulsar
PSR~J0737$-$3039 \citep{Kramer:2016kwa}. Bounds on the graviton mass from
individual pulsars are illustrated in Fig.~\ref{fig:bounds}. Because the
graviton mass is a universal quantity in these tests, a combination is
possible. Combining the bounds with a Bayesian analysis,
\citet{Shao:2020fka} gives,
\begin{equation}
  m_{g} < 2 \times 10^{-28} \, \mathrm{eV} / c^{2} \,, \quad \mbox{(95\%
  {C.L.})} \,, \label{eq:cubic:Galileon:limit}
\end{equation}
when a uniform prior on $\ln m_g$ for $m_g \in \big( 10^{-29}, 10^{-27}
\big) \, \mathrm{eV} / c^{2} $ is adopted. The bound
(\ref{eq:cubic:Galileon:limit}) is specific to the cubic Galileon model
(\ref{eq:cubic}).

\section{Tests via pulsar timing arrays}\label{sec:cubic}

Since the early 2000s, dozens of well-timed millisecond pulsars have been
used to detect GWs at the nano-hertz frequency. A stochastic GW background
imprints an angular-dependent correlation in pulsar timing residuals for
pulsars distributed across our Milky Way \citep{Hellings:1983fr}.
Therefore, it is possible to use PTAs to extract the angular correlation in
pulse signals and derive the physical information of GWs
\citep{Hobbs:2009yy}.

When GWs are not described by the GR, the Hellings-Downs correlation is
changed accordingly. For a massive graviton satisfying the
Lorentz-invariant dispersion relation,
\begin{equation}
  E^{2}=p^{2} c^{2}+m_g^{2} c^{4} \,,
\end{equation}
\citet{Lee:2010cg} derived the changes to the canonical case. Most
importantly, the shift in the frequency of pulsar timing radio signals, by
a monochromatic plane GW with a frequency $\omega_g$ and a wave vector
${\bm{k}}_{\mathrm{g}}$, is given by,
\begin{equation}
  \frac{\Delta \omega(t)}{\omega}= \sum_{ij} -\frac{\hat{{n}}^{i}
  \hat{{n}}^{j} \left[h_{i j}(t, \bm{0})-h_{i j}\left(t-{|\bm{D}|},
  \bm{D}\right)\right] }{2\left[1+ ({\bm{k}}_{\mathrm{g}} /
  \omega_{\mathrm{g}}) \cdot \hat{\bm{n}}\right]} \,,
\end{equation}
where $\hat{\bm{n}}$ is the direction from the Earth (at location $\bm{r} =
\bm{0}$) to the pulsar (at location $\bm{r} = \bm{D}$), and $h_{i j}$ is
the strain of GWs \citep{Lee:2010cg}. Then the change in the timing
residual is obtained via $R(t)=\int_{0}^{t} {\Delta \omega(\tau)}/{\omega}
{\rm d} \tau $. In addition to the modification of the timing residuals,
the presence of a massive graviton removes all GW radiating powers at
frequencies below the cutoff frequency which is defined by the mass of
graviton.

Extensive simulation shows that, with a large sample of stable pulsars, the
mass of the graviton will be bound to about $10^{-22} \, \mathrm{eV} /
c^{2}$ with realistic PTAs \citep{Lee:2010cg}. Inclusion of extra
polarization modes \citep{Eardley:1974nw}, other than the canonical
``plus'' and ``cross'' ones, will not change much of the projected limit on
$m_g$ \citep{Lee:2014awa}.

\section{Summary}\label{sec:disc}

When there are apparent conflicts between GR and quantum field theory, it
is motived to look for new physics beyond the standard paradigm. The
gravity might be holding the key to a breakthrough in the field
\citep{Will:2018bme, Berti:2015itd}. Various experimental examination was
carried out to different catalogs of alternative gravity theories,
including the massive gravity \citep{deRham:2014zqa,deRham:2016nuf}.

This contribution overviews three methods in the field that uses the
precision timing of radio pulsars to bound the graviton mass: two of them
use the gravitational radiation backreaction in binary pulsars
\citep{Finn:2001qi, Miao:2019nhf, deRham:2012fw, Shao:2020fka}, while the
other one uses the angular-correlation curve in PTAs \citep{Lee:2010cg,
Lee:2014awa}. The above derived/projected bounds on the graviton mass from
radio pulsars, together with those from other experiments---as
comprehensively reviewed in \citet{deRham:2016nuf}---are all of value to
the field of experimental gravity, since they are based on different
assumptions about the (unknown) theory of massive gravity. They have
different powers at probing the hypothesis of a massive graviton.
Comparison between them is meaningful only when the underlying theory and
assumptions are made clear \citep{deRham:2016nuf}.

The tests are to be sharpened to a new level with new instruments and
continued observations, in particular with the demonstrated capability of
the South African MeerKAT radio telescope \citep{Bailes:2018azh} and the
Chinese Five-hundred-meter Aperture Spherical Telescope
\citep[FAST;][]{Jiang:2019rnj, Lu:2019gsr}, and ultimately with the Square
Kilometre Array \citep[SKA;][]{Kramer:2004hd, Bull:2018lat, Shao:2014wja}.
Radio pulsars will continue to provide interesting gravity tests, in
complement to tests from other fields \citep{Wex:2014nva, Yunes:2016jcc,
Shao:2017gwu, Sathyaprakash:2019yqt}.


\section*{Acknowledgments}

It is my pleasure to thank the Organizing Committee of IWARA\,2020 for the
invitation, and Kejia Lee, Xueli Miao, Norbert Wex, and Shuang-Yong Zhou
for discussions.
This work was supported by the \fundingAgency{National Natural Science
Foundation of China} under Grant Nos. \fundingNumber{11991053, 11975027,
11721303}, the Young Elite Scientists Sponsorship Program by the
\fundingAgency{China Association for Science and Technology} under the
Grant No. \fundingNumber{2018QNRC001}, and the \fundingAgency{Max Planck
Society} through the \fundingNumber{Max Planck Partner Group}. It was
partially supported by the Strategic Priority Research Program of the
\fundingAgency{Chinese Academy of Sciences} under the Grant No.
\fundingNumber{XDB23010200}, and the High-performance Computing Platform of
Peking University.

\subsection*{Conflict of interest}

The author declares no potential conflict of interests.

\bibliography{refs}%


\end{document}